\documentclass[aps,pra,footinbib,twocolumn,superscriptaddress]{revtex4-1} 
\usepackage[colorlinks=true,citecolor=black,urlcolor=black,linkcolor=black]{hyperref} 
\usepackage{dsfont}
\usepackage{amsmath}
\usepackage{graphicx}
\usepackage{multirow}
\usepackage[latin1]{inputenc}
\usepackage{ulem} 
\usepackage{units} 
\usepackage{version} 
\usepackage{color}
\usepackage{colortbl}
\usepackage{xcolor}  

\makeindex
\begin{document}

\title{Double light-cone dynamics establish thermal states in integrable 1D Bose gases}

\author{T. Langen}
\email{t.langen@physik.uni-stuttgart.de}  

\affiliation{5. Physikalisches Institut and Center for Integrated Quantum Science and Technology (IQST), Universit\"at Stuttgart, Pfaffenwaldring 57, 70569 Stuttgart, Germany}

\author{T. Schweigler}
\affiliation{Vienna Center for Quantum Science and Technology, Atominstitut, TU Wien, Stadionallee 2, 1020 Vienna, Austria}

\author{E. Demler}
\affiliation{Physics Department, Harvard University, Cambridge, Massachusetts 02138, USA}

\author{J. Schmiedmayer}
\affiliation{Vienna Center for Quantum Science and Technology, Atominstitut, TU Wien, Stadionallee 2, 1020 Vienna, Austria}

\begin{abstract} 
We theoretically investigate the non-equilibrium dynamics in a quenched pair of 1D Bose gases with density imbalance. We describe the system using its low-energy effective theory, the Luttinger liquid model. In this framework the system shows strictly integrable relaxation dynamics via dephasing of its approximate many-body eigenstates. In the balanced case, this leads to the well-known light-cone-like establishment of a prethermalized state, which can be described by a generalized Gibbs ensemble. In the imbalanced case the integrable dephasing leads to a state that, counter-intuitively, closely resembles a thermal equilibrium state. The approach to this state is characterized by two separate light-cone dynamics with distinct characteristic velocities. This behavior is rooted in the fact that in the imbalanced case observables are not \textit{aligned} with the conserved quantities of the integrable system. We discuss a concrete experimental realization to study this effect using matterwave interferometry and many-body revivals on an atom chip. 
\end{abstract}

\date{\today}

\maketitle

\section{INTRODUCTION}
Non-equilibrium dynamics of isolated quantum systems play a central role in many fields of physics~\cite{Polkovnikov11}. An important question in this context is whether the unitary evolution can lead to the emergence of thermal properties. For example, the eigenstate thermalization hypothesis conjectures that dephasing can lead to thermalization in systems with a chaotic classical limit~\cite{Deutsch91,Srednicki94,Rigol08,Kaufman2016}. On the other hand, integrable or many-body localized systems are expected not to thermalize at all~\cite{Rigol09}, but instead relax to generalized thermodynamical ensembles~\cite{Jaynes57b,Rigol07,Vosk2013,Langen2015}. An important role in both cases is played by the signal propagation during the non-equilibrium dynamics. It has been shown for many systems~\cite{Lieb72,Cheneau12,Langen13b}  that this propagation follows a \textit{light-cone-like} linear evolution in time with a characteristic velocity. This behavior has important consequences for the growth of entanglement in such systems~\cite{Eisert2010,Kaufman2016}.

In this manuscript, we study the dynamics in a pair of bosonic 1D quantum gases with number imbalance using the Luttinger liquid formalism. The dynamics of such \textit{quantum wires} has recently been studied in great detail using atom chips~\cite{Gring12,Langen2015,Langen13b,Jacqmin11,Schweigler17} or in optical lattices~\cite{Widera08,Haller09,Kinoshita04,Kinoshita06,Paredes04}. In particular, 1D Bose gases have been established as a prime experimental realization of a nearly integrable system with strongly suppressed thermalization~\cite{Kinoshita06,Gring12,Rigol09}. So far, the consequences of this integrable behavior have mainly been studied for individual gases or sets of nearly identical gases. Here, we show that for imbalanced pairs of gases, i.e. gases that differ in their mean density, integrable dephasing can establish a state that closely resembles thermal equilibrium. The dynamics towards this state are found to be exceptionally rich, including a metastable thermal-like state described by a generalized Gibbs ensemble and two distinct light-cone dynamics, each exhibiting their individual characteristic velocity. Beyond the fundamental interest in thermalization dynamics, our study is of high relevance for ongoing experiments, where density imbalances are often unavoidable and can thus fundamentally affect the interpretation of the results.

\begin{figure}[tb]
	\centering
		\includegraphics[width=0.95\columnwidth]{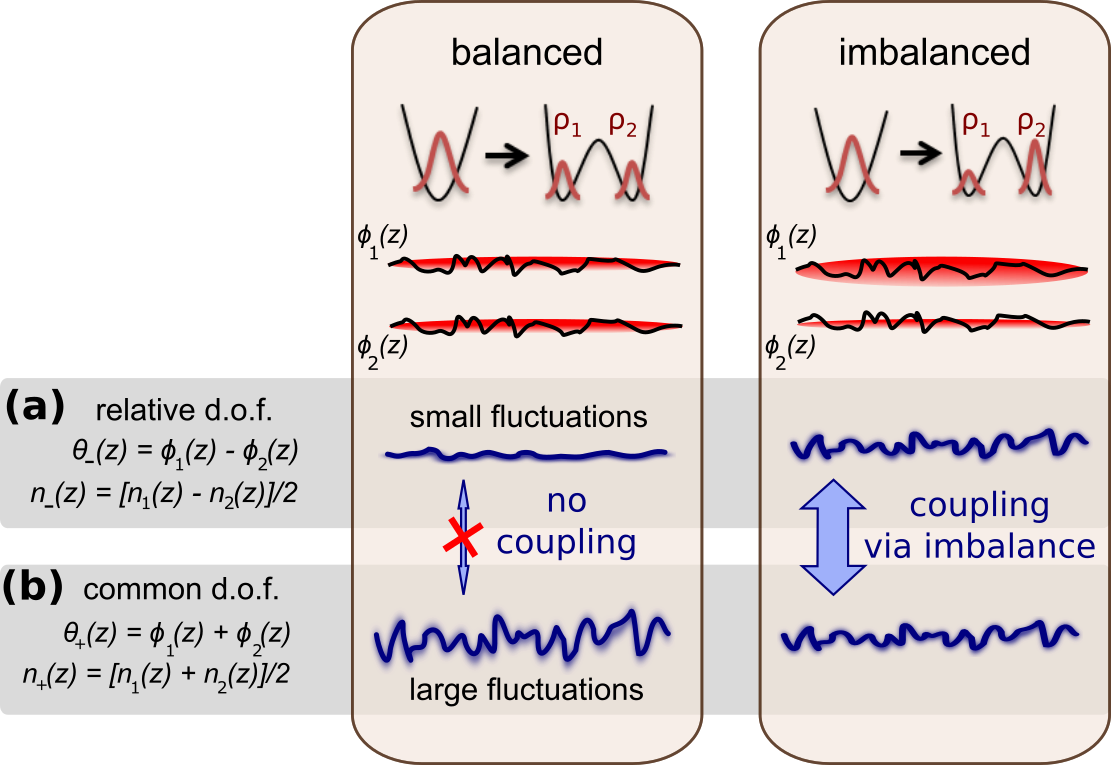}
	\caption{\textbf{Relaxation dynamics in a pair of quantum wires.} A phase-fluctuating 1D Bose gas in thermal equilibrium is coherently split into a pair of gases using a double well potential. This leads to almost identical phase profiles $\phi_1(z)$ and $\phi_2(z)$ (indicated by the solid black lines) for the two resulting gases and to a binomial distribution of atoms between them. We study here the relaxation of this highly-correlated state. If the two gases after the split have the same densities $\rho_1=\rho_2$, there is no coupling between the anti-symmetric or \textit{relative} degrees of freedom (d.o.f.), which contain only  small fluctuations after the quench, and the symmetric or \textit{common} degrees of freedom of the total system, which contain large fluctuations after the quench. In the dynamics, these fluctuations of the relative and common d.o.f. reach thermal-like steady states with different temperatures. If, however, the two gases after the split have different densities, relative and common degrees of freedom are coupled and their fluctuations equilibrate over a second, much slower timescale. }
	\label{fig:figure1}
\end{figure}

\section{MODEL}
In the following we consider an initial 1D Bose gas in thermal equilibrium that is coherently split into two gases (Fig.~\ref{fig:figure1}). This scenario is motivated by experiments with 1D Bose gases on atom chips, which have recently been established as an important model system to study non-equilibrium dynamics~\cite{Gring12,Langen2015,Langen13b,langen2015thesis,Rauer17}. In these experiments gases are confined in two radial directions using a strong potential characterized by the trapping frequency $\omega_\perp$, such that $\mu,k_B T <\hbar\omega_\perp$ and they behave effectively one-dimensional. Here, $\mu$ is their chemical potential and $T$ their temperature. 

The effective low-energy description for each of the gases is given by the Luttinger liquid (LL) Hamiltonian
\begin{equation}
\hat H_i = \int dz\left[\frac{\hbar^2 \rho_i}{2m} \left(\nabla\hat\phi_i(z)\right)^2 + \frac{g}{2}\left(\hat n_i(z)\right)^2\right].\label{eqn:llhamiltonian}
\end{equation}
Here, the index $i=1,2$ labels the individual gases with mean densities $\rho_i$ independent of $z$ (homogeneous case). We denote the density fluctuations around this mean densities by $\hat n_i(z)$, the phase fluctuations by $\hat\phi_i(z)$. Phase and density fluctuations represent conjugate variables. The 1D interaction strength is characterized by $g=2\hbar a_s \omega_\perp$, with the 3D scattering length $a_s$. 

The resulting total Hamiltonian of the system is given by $\hat H = \sum_{i=1,2}\hat H_i$. To analyze this Hamiltonian we transfer it into a basis formed by the symmetric and anti-symmetric superpositions of the individual fluctuations:
\begin{align}
\begin{split}
\hat n_{\pm}(z)&=[\hat n_1(z) \pm \hat n_2(z)]/2 \\
\hat\theta_{\pm}(z)&=[\hat\phi_1(z) \pm \hat\phi_2(z)] \label{eq:transf_sym_anti}
\end{split}
\end{align}
This basis transformation allows us to directly connect our results to experiments, which probe the relative phase $\hat\theta_-(z)$ (i.e. the anti-symmetric degrees of freedom) between the gases through matterwave interference~\cite{Schumm05,Langen13b,Gring12}.

After this basis transformation we can write the total Hamiltonian as~\cite{Kitagawa11}
\begin{align}
\hat H = \hat H_- + \hat H_+ + \hat H_\mathrm{c},
\end{align}
where  $\hat H_-$ and $\hat H_+$ are again LL Hamiltonians
\begin{equation}
\hat H_{\pm} = \int dz\left[\frac{\hbar^2 (\rho_1+\rho_2)}{8 m} \left(\nabla\hat\theta_{\pm}(z)\right)^2 + g \left(\hat n_{\pm}(z)\right)^2\right].\label{eqn:llhamiltonian2}
\end{equation}
We note the change in pre-factors, i.e. the factor $1/2$ that appears in addition to the mean density $(\rho_1+\rho_2)/2$  in the first term and the appearance of $g$ instead of $g/2$ in the second term of $H_\pm$, as compared to the original Luttinger Hamiltonians decribing the individual gases~(Eq.~\ref{eqn:llhamiltonian}). These changes arise due to our specific definition of the symmetric and anti-symmetric degrees of freedom in Eq.~\ref{eq:transf_sym_anti}. Coupling between the new degrees of freedom is mediated by 
\begin{align}
\hat H_\mathrm{c} = \int dz\left[\frac{\hbar^2(\rho_1-\rho_2)}{4m}\left(\nabla\hat\theta_+(z)\nabla\hat\theta_-(z)\right)\right].
\end{align}
All these Hamiltonians are still quadratic and integrable and we thus do not expect any thermalization. 

At this point there are two distinct cases to study. If the mean densities  $\rho_i$ of the two wires are identical, i.e. there is \textit{no imbalance} $\Delta = (\rho_1-\rho_2)/(\rho_1+\rho_2)=0$, we have $\hat H_\mathrm{mix}=0$ and symmetric and anti-symmetric degrees of freedom decouple. In this case, fluctuations in the individual gases evolve in exactly the same way. However, if the mean densities of the two wires are different ($\Delta \neq 0$), we have $\hat H_\mathrm{c} \neq 0$, which leads to a coupling between symmetric and anti-symmetric degrees of freedom. In the latter case, the fluctuations in the two condensates evolve differently causing a relative dephasing of the two gases over time.

We note in passing that this scenario holds promise as an experimental platform for spin-charge physics within the Luttinger liquid framework~\cite{Giamarchi04,Schmidt10}, with $\hat H_\pm$ describing the spin and charge degrees of freedom, respectively. Details of a possible, fully tunable experimental realization are outlined in the Appendix. 

Previously, the corresponding dynamics were obtained by diagonalizing $\hat H_-$~\cite{Kitagawa10,Kitagawa11,Geiger13,Langen13}. However, this is not sufficient to capture the full dynamics if $\hat H_\mathrm{mix}\neq 0$. Instead we diagonalize $\hat H_1$ and $\hat H_2$ individually. This is possible, as their excitations are conserved independently of the imbalance. 
Assuming periodic boundary conditions for the gases of length $L$, we can expand the phase and density fluctuations into their Fourier components
\begin{align}
\begin{split}
\hat \phi_k &= \frac{1}{\sqrt{L}} \int_{0}^{L} dz \ e^{- i k z} \ \hat \phi(z) \\
\hat n_k &= \frac{1}{\sqrt{L}} \int_{0}^{L} dz \ e^{- i k z} \ \hat n(z) 
\end{split}
\end{align}
leading to
\begin{equation}
\hat H_i = \sum_k \frac{\hbar^2 k^2}{2m}\rho_i \hat\phi_{i,k}^\dagger\hat\phi_{i,k}+\frac{g}{2}\hat n_{i,k}^\dagger\hat n_{i,k}.
\label{eq:HLLk}
\end{equation}

Using this Hamiltonian to solve the equation of motion for the phase operator we find
\begin{align}
\hat \phi_{i,k}(t)=\,& \hat \phi_{i,k}(t=0)\cos(c_i kt)\nonumber\\&-\frac{g}{\hbar c_i k}\hat n_{i,k}(t=0)\sin(c_i kt). \label{eq:operator_evolution}
\end{align}
Here, the speed of sound in the individual gases is given by $c_i = \sqrt{g \rho_i/ m}$, while 
$\phi_{i,k}(t=0)$ and $\hat n_{i,k}(t=0)$ denote the initial values for the phase and density fluctuations, respectively. We can transfer this result into the familiar symmetric/anti-symmetric basis using $\hat \theta_{\pm,k} = \hat\phi_{1,k}\pm\hat \phi_{2,k}$. 

Our aim is to investigate the dynamics after a single 1D gas in thermal equilibrium (with temperature $T_\mathrm{in}$) is split into two parts. The Hamiltonian of this initial gas is of the form given by Eq.~\ref{eq:HLLk} with 1D density $\rho = \rho_1 + \rho_2$. The thermal expectation values for the second moments in classical field approximation are therefore
\begin{align}
\begin{split}
\langle \hat{n}_{k} \ \hat{n}_{k'}  \rangle_{\mathrm{th}} &= \frac{k_B T_\mathrm{in}}{g_\mathrm{in}} \cdot \delta_{k,-k'} \\
\langle \hat{\phi}_{k} \ \hat{\phi}_{k'}  \rangle_{\mathrm{th}} &= \frac{m k_B T_\mathrm{in}}{\hbar^2 k^2 (\rho_1 + \rho_2)}  \cdot \delta_{k,-k'} \\
\langle \hat{n}_{k} \ \hat{\phi}_{k'}  \rangle_{\mathrm{th}} &= 0,
\end{split}
\end{align} 
where we have used $\hat{\phi}_{k}^\dagger = \hat{\phi}_{-k}$ and $\hat{n}_{k}^\dagger = \hat{n}_{-k}$. All first moments vanish. Note that experimentally the splitting often results in a change in radial trapping frequency, such that the gases before and after splitting will differ in their 1D interaction strength $g\sim\omega_\perp$. In the following, we will denote the interaction strength before splitting with $g_\mathrm{in}$ and after splitting with $g_\mathrm{f}$ to include this possibility in our calculations. 

In the limit of fast splitting no information can propagate along the axis of the gases and we can assume that they both have the same phase profile after the splitting, which is identical to the profile of the initial gas. Moreover, we want to assume that in this case the probability for each atom to go into well 1 is $\rho_1/(\rho_1 + \rho_2)$ independently from where the other atoms go (binomial splitting)~\cite{Kitagawa11}. With these assumptions, we find for the fluctuations of the individual gases right after splitting
\begin{widetext}
\begin{align}
\langle \hat{n}_{i,k} \ \hat{n}_{j,k'}  \rangle (t=0) &=  \left[ (\rho_1 + \rho_2) \cdot \frac{\rho_1 \rho_2}{(\rho_1 + \rho_2)^2} \cdot  (2 \delta_{i,j} - 1)  + \frac{k_B T_\mathrm{in}}{g_\mathrm{in}} \cdot \frac{\rho_i \rho_j}{(\rho_1 + \rho_2)^2}\right] \cdot \delta_{k,-k'} \label{eq:dens_init}\\
\langle \hat{\phi}_{i,k} \ \hat{\phi}_{j,k'}  \rangle (t=0) &= \langle \hat{\phi}_{k} \ \hat{\phi}_{k'}  \rangle_{\mathrm{th}} = \frac{m k_B T_\mathrm{in}}{\hbar^2 k^2 (\rho_1 + \rho_2)}  \cdot \delta_{k,-k'} \label{eq:ph_init} \\
\langle \hat{\phi}_{i,k} \ \hat{n}_{j,k'}  \rangle(t=0) &= 0. \label{eq:mix_init}
\end{align}
\end{widetext}
The first term in Eq.~\ref{eq:dens_init} represents the shot noise from the binomial splitting process, which is anti-correlated, as expressed by the factor $2 \delta_{i,j} - 1$. The second term in Eq.~\ref{eq:dens_init}, as well as Eq.~\ref{eq:ph_init}, stem from the thermal fluctuations of the initial condensate, and describe correlated fluctuations. Again, all first moments vanish. Assuming Gaussian fluctuations, the second moments are sufficient to fully describe the system. This assumption is justified for long enough length scales containing a large number of particles. Note that the same assumption has to be made for the validity of the Luttinger liquid model and, also, typically only such length scales are accessible in experiments.

\section{RESULTS}
We can now investigate the dynamics by combining Eqs.~\ref{eq:dens_init}-\ref{eq:mix_init} with Eq.~\ref{eq:operator_evolution}. Assuming only a small imbalance ($\Delta\ll 1$) between the two gases, we obtain the following approximation for the time evolution of the relative phase variance
\begin{align}
\langle |\theta_{-,k}(t)|^2\rangle&= \frac{m k_B T_\mathrm{in}}{\hbar^2 k^2 (\rho_1 + \rho_2)}\sin^2(c_- kt) \nonumber\\&\times \left[2+\frac{g_\mathrm{f}}{g_\mathrm{in}}-\left(2-\frac{g_\mathrm{f}}{g_\mathrm{in}}\right)\cos(2c_+kt)\right]\nonumber\\
	&+\frac{2mg_\mathrm{f}}{\hbar^2k^2}\sin^2(c_+kt)\cos^2(c_- kt).\label{eqn:phasevariance}
\end{align}
with the average velocity $ c_+=(c_1+c_2)/2$ and and the velocity difference $c_-=(c_1-c_2)/2$. 

In the following discussion of the dephasing dynamics we first focus on the case when the length scales under consideration are much smaller then the system size (large/infinite system limit). Finite system sizes and the occurrence of revivals are discussed in section~\ref{sec:revivals}.

If there is no imbalance (i.e. $\Delta = 0$ and  $c_-=0$, $c_+ = c_1 = c_2 = c$) the terms proportional to $\sin^2(c_- kt)$ vanish and we obtain the well known dephasing dynamics
\begin{equation}
\langle |\theta_{-,k}|^2\rangle = \frac{2mg_\mathrm{f}}{\hbar^2k^2}\sin^2(c k t) \label{eqn:phasevariance_no_im},
\end{equation}
where thermal correlations are established with a light-cone~\cite{Langen13}. In this process thermal correlations instantaneously emerge locally within a certain horizon,
while they remain non-thermal outside of the horizon. This horizon spreads through the system with a characteristic velocity that is given by $c=c_1=c_2$~\cite{Langen13,Langen13b,Geiger13}. One can see this from Eq.~\ref{eqn:phasevariance_no_im} by realizing that at a certain time $t$ all modes down to a lower bound given by $2 c t \cdot  k_\mathrm{lower} = 2 \pi$ have dephased (note that the factor 2 comes from the square of the sine). The bound $k_\mathrm{lower}$ therefore corresponds to the length-scale $2 c t$. The result of the dephasing dynamics is a prethermalized state with a temperature
\begin{equation}
T_\mathrm{eff}^{(-)}= \frac{g_\mathrm{f} (\rho_1 + \rho_2)}{4 k_B}.\label{eqn:teff}
\end{equation}
This temperature can be identified directly from Eq.~\ref{eqn:phasevariance} in the dephased limit, i.e. by averaging over $k t$ and comparing to the result for a pair of gases (each with 1D density $(\rho_1 + \rho_2)/2$) in thermal equilibrium $\langle |\theta_k|^2\rangle_\mathrm{th}=\frac{4mk_B T}{\hbar^2 k^2 (\rho_1 + \rho_2)}$ (classical field approximation). It corresponds to the energy $k_B T_\mathrm{eff}^{(-)}$ that is added to the relative degrees of freedom during the splitting quench~\cite{Kitagawa10,Gring12}. 

The corresponding expression to Eq.~\ref{eqn:phasevariance} for the common phase variance $\langle |\theta_{+,k}(t)|^2\rangle$ can be obtained in exactly the same way (see Appendix). From this, one observes that the symmetric degrees of freedom exhibit the temperature 
\begin{equation}
T_\mathrm{eff}^{(+)}=  \frac{T_\mathrm{in}}{2}\left(1 + \frac{1}{2} \cdot \frac{g_\mathrm{f}}{g_\mathrm{in}}\right),\label{eqn:tcommon}
\end{equation}
which is coming from the initial thermal fluctuations. However, the initial temperature is decreased through an interaction quench (second term in the brackets). In the splitting not only the density but also the density-fluctuations are halved~\cite{Rauer16,Grisins16,Johnson17}. As they enter quadratically in the Hamiltonian, this leads to a decrease of a factor $2$ in energy, which can be further modified by the aforementioned change in the interaction constant $g$.

While the individual correlation functions of relative and common degrees of freedom are thus thermal, the state of the total system is non-thermal and has to be described by a generalized Gibbs ensemble with the two temperatures $T_\mathrm{eff}^{(\pm)}$, respective~\cite{Gring12,Langen2015}. 

\textit{With imbalance} ($|\Delta|>0$) the behavior of the dephasing dynamics changes significantly. The aforementioned light-cone dynamics to the prethermalized state still proceeds with the average velocity $c_+$. In addition, examining the additional terms in Eq.~\ref{eqn:phasevariance} we identify a second dephasing timescale characterized by the slower velocity $c_-$. After complete dephasing (with the fast as well as the slow velocity), we end up with a second thermal-like state. Following the same procedures as before we can identify the temperature of this state to be identical for both relative and common degrees of freedom and given by~\footnote{Note that this temperature is obtained by comparison with the thermal equilibrium of two gases of equal density $(\rho_1 + \rho_2)/2$. When comparing to two gases with unequal density $\rho_1$ and $\rho_2$, the effective temperature is given by $4\rho_1\rho_2 /(\rho_1 + \rho_2)^2\times T_f^\pm$}
\begin{align}
T_\mathrm{f}^{(\pm)}=\frac{T_\mathrm{in}}{4}\left( 1+ \frac{1}{2}\frac{g_\mathrm{f}}{g_\mathrm{in}}\right) + \frac{T_\mathrm{eff}}{2}.\label{eqn:temperature}
\end{align}
Again, this result can be interpreted intuitively in terms of the corresponding energies $k_B T_\mathrm{f}^{(\pm)}$. The first term corresponds to half the energy that is initially contained in the common degrees of freedom (Eq.~\ref{eqn:tcommon}), the second term to half the energy introduced to the relative degrees of freedom during the quench (Eq.~\ref{eqn:teff}). Eq.~\ref{eqn:temperature} hence describes an equipartition of energy that is dynamically established by the coupling term $H_c$. 

Note that Eq.~\ref{eqn:temperature} remains true, even without the assumption of small imbalance, which was used to obtain Eq.~\ref{eqn:phasevariance}. For typical parameters in atomchip microtraps the change in confinement leads to $\frac{g_\mathrm{f}}{g_\mathrm{in}} \approx 1/\sqrt{2}$  and we find $T_\mathrm{f}\approx\frac{T_\mathrm{in}}{3}+\frac{T_\mathrm{eff}}{2}$. 

\begin{figure}[htb]
	\centering
	\includegraphics[width=0.44\textwidth]{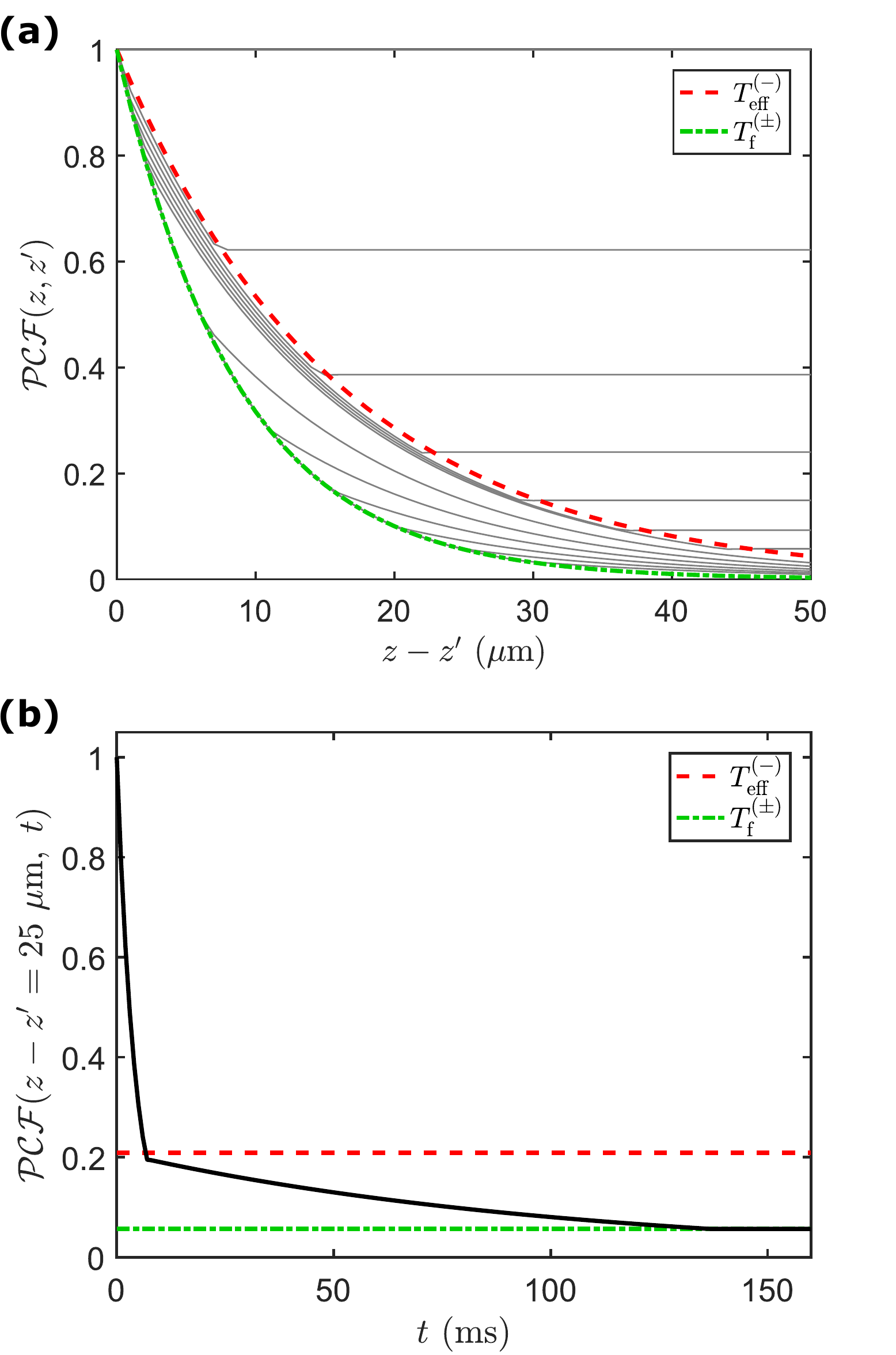}
	\caption{\textbf{Dynamics of the two-point phase correlation function in the theoretical model.} (a) The two-point phase correlation function reveals a fast light-cone-like decay with the average velocity $c_+$ to the prethermalized state (red line), followed by a much slower second light-cone-like decay with the difference velocity $c_-$ to the final relaxed state (green line). Evolution times increase in steps of $2\,$ms from top to bottom until the prethermalized state is reached, followed by steps of $25\,$ms in the approach to the final state. Parameters are $a_s = 5.24\,$nm and $m=1.44\times10^{-25}\,$kg for ${}^{87}$Rb, $\omega_{\perp,\mathrm{in}}=2\pi\times 2.1\,$kHz, $\rho_1+\rho_2 =\,100\,\mu\mathrm{m}^{-1}$, corresponding to an initial Luttinger parameter $K=\hbar\pi\sqrt{(\rho_1+\rho_2)/mg}\sim 74$. The imbalance is $\Delta = (\rho_1-\rho_2)/(\rho_1+\rho_2)=0.1\,$ and $T_\mathrm{in}=70\,$nK. The trap frequency after splitting is  $\omega_{\perp,\mathrm{f}}=2\pi\times 1.4\,$kHz. (b) Phase correlation function evaluated for a distance of $z - z' = 25\,\mu$m $\sim 105 \times \xi_h$ as a function of time, highlighting the distinct timescales of the relaxation to the prethermalized and final states, respectively. Here, $\xi_h=\hbar/mc$ is the healing length.}
	\label{fig:figure2}
\end{figure}

To visualize the corresponding dephasing dynamics leading to this equipartition in detail, we calculate the two-point phase correlation function~\cite{Langen13b} 
\begin{equation}
\mathcal{PCF}(z,z^\prime) = \langle \cos(\hat\theta(z)-\hat\theta(z^\prime))\rangle.
\end{equation}
This function measures the correlation between the relative phases $\theta(z)$ at two arbitrary points $z$ and $z^\prime$ along the length of the system and can directly be measured in the experiments~\cite{Langen13b}. 

As discussed in the previous section, the initial fluctuations in our model are Gaussian and of course remain so during the evolution with the quadratic Hamiltonian. Therefore, the phase correlation function can be rewritten in the form ${\cal PCF}(z,z^\prime) = e^{-\frac{1}{2}\langle{[\hat\theta(z)-\hat\theta(z^\prime)]^2}\rangle}$. In the limit of an infinitely large system this leads to
\begin{align}
\mathcal{PCF}(z,z^\prime)=&\exp\bigg[-\int_0^\infty\frac{dk}{\pi}\left\langle|\theta_{-,k}(t)|^2\right\rangle\nonumber\\
&\times (1-\cos k(z-z'))\bigg].\label{eqn:PCF}
\end{align}
In Fig.~\ref{fig:figure2} we plot Eq.~\ref{eqn:PCF} for increasing evolution times, revealing a double light-cone. First, the system relaxes to the prethermalized state with exponentially decaying (thermal) correlations. For longer evolution times, the system relaxes further to the second thermal-like steady state. As in the previous light-cone-like relaxation to the prethermalized state, the system reaches this new final state for a given time only up to a certain horizon, but then follows a different shape beyond that point. The position of this horizon moves with a second characteristic velocity that is given by the (typically small) velocity difference $c_-$ of the individual gases. 

While both symmetric and anti-symmetric degrees of freedom reach a thermal-like state with temperature $T_\mathrm{f}^{(\pm)}$, the complete system still differs from the thermal equilibrium of two condensates with equal density $(\rho_1 + \rho_2)/2$ in the aspect that cross-correlations $\langle\theta_{+,k}\theta_{-,k}\rangle$ between symmetric and anti-symmetric degrees of freedom do not vanish (see Appendix). Note that for the thermal equilibrium of two gases with unequal densities, the cross-correlations between common and relative degrees of freedom also do not vanish. However, they are still of different magnitude than in the completely dephased case.

Another way to discuss the question of whether the system dephases to thermal equilibrium is to have a look at the quantities for the individual gases. These could e.g. be studied in experiments using density fluctuations in time of flight~\cite{Manz10} or by probing the density fluctuations in situ~\cite{Esteve06}. The phase variance of the individual gases is given by
\begin{align}
\begin{split}
\langle|\phi_{i,k}|^2\rangle = & \frac{m k_B}{\hbar^2 k^2 \rho_i}\bigg[ T_\mathrm{in}\frac{\rho_i}{\rho_1+ \rho_2}\\ &\left(\cos^2(c_ikt)+  \frac{g_\mathrm{f}}{g_\mathrm{in}}\frac{\rho_i}{\rho_1 + \rho_2}\sin^2(c_ikt)\right) \\ &+\frac{\rho_i g_\mathrm{f}}{k_B} \left(1-\frac{\rho_i}{\rho_1 + \rho_2}\right)\sin^2(c_i kt)\bigg]
\label{eqn:singlegases_phvar_evo}
\end{split}
\end{align}
which describes a relaxation towards a temperature
\begin{align}
\begin{split}
T^{(i)}_\mathrm{f}&=\frac{T_\mathrm{in}}{2}\frac{\rho_i}{\rho_1 + \rho_2}\left(1+\frac{g_\mathrm{f}}{g_\mathrm{in}}\frac{\rho_i}{\rho_1+\rho_2}\right) \\
&+\frac{\rho_i g_\mathrm{f}}{2k_B}\left(1-\frac{\rho_i}{\rho_1 + \rho_2}\right).
\label{eqn:singlegases}
\end{split}
\end{align}
This expression is different from the results for the symmetric/anti-symmetric basis, highlighting how the observed dynamics and, in particular, also their timescales, are indeed intimately connected to the choice of observable. 

In detail, the time scale for the dynamics within a single gas is, as expected, given by their speed of sound $c_i$. However, the cross-correlations of the form $\langle \phi_{1,k} \phi_{2,k} \rangle$ dephase to zero with the slow velocity $c_-$ (see Appendix). After complete dephasing we therefore end up with two independent gases, which independently appear to be in thermal equilibrium with their respective temperatures $T^{(i)}_\mathrm{f}$.  However, for all the dynamics described only dephasing and no true thermalization has taken place. The intuitive reason for this complex behavior is that the two imbalanced gases are non-identical and dephase with respect to each other. Therefore, their individual excitations are still conserved, but the symmetric and anti-symmetric modes are no longer connected to these conserved quantities. 

As an example, for the parameters used in Fig.~\ref{fig:figure2} ($\Delta=0.1\,$, ${}^{87}$Rb atoms with $T_\mathrm{in}=70\,$nK and $\rho_1+\rho_2 =\,100\,\mu\mathrm{m}^{-1}$) these final temperatures are $T^{(1)}_\mathrm{f}=43.8\,$nK and $T^{(2)}_\mathrm{f}=35.1\,$nK. Note that in the limit of vanishing imbalance we have $\rho_i/(\rho_1+\rho_2)\sim1/2$ and the difference between the $T^{(i)}_\mathrm{f}$ tends to zero and they approach the final temperature of the symmetric and anti-symmetric degrees of freedom given by Eq.~\ref{eqn:temperature}. Also, already in the approach of this limit the small difference in final temperatures can be challenging to measure in an experiment. In both cases the system would thus appear completely thermalized independent of the choice of basis, with symmetric, anti-symmetric and individual degrees of freedom all exhibiting the same temperature. However, with imbalance going to zero, this approach of the final temperature would become infinitely slow. 

\begin{figure*}
	\centering
		\includegraphics[width=0.98\textwidth]{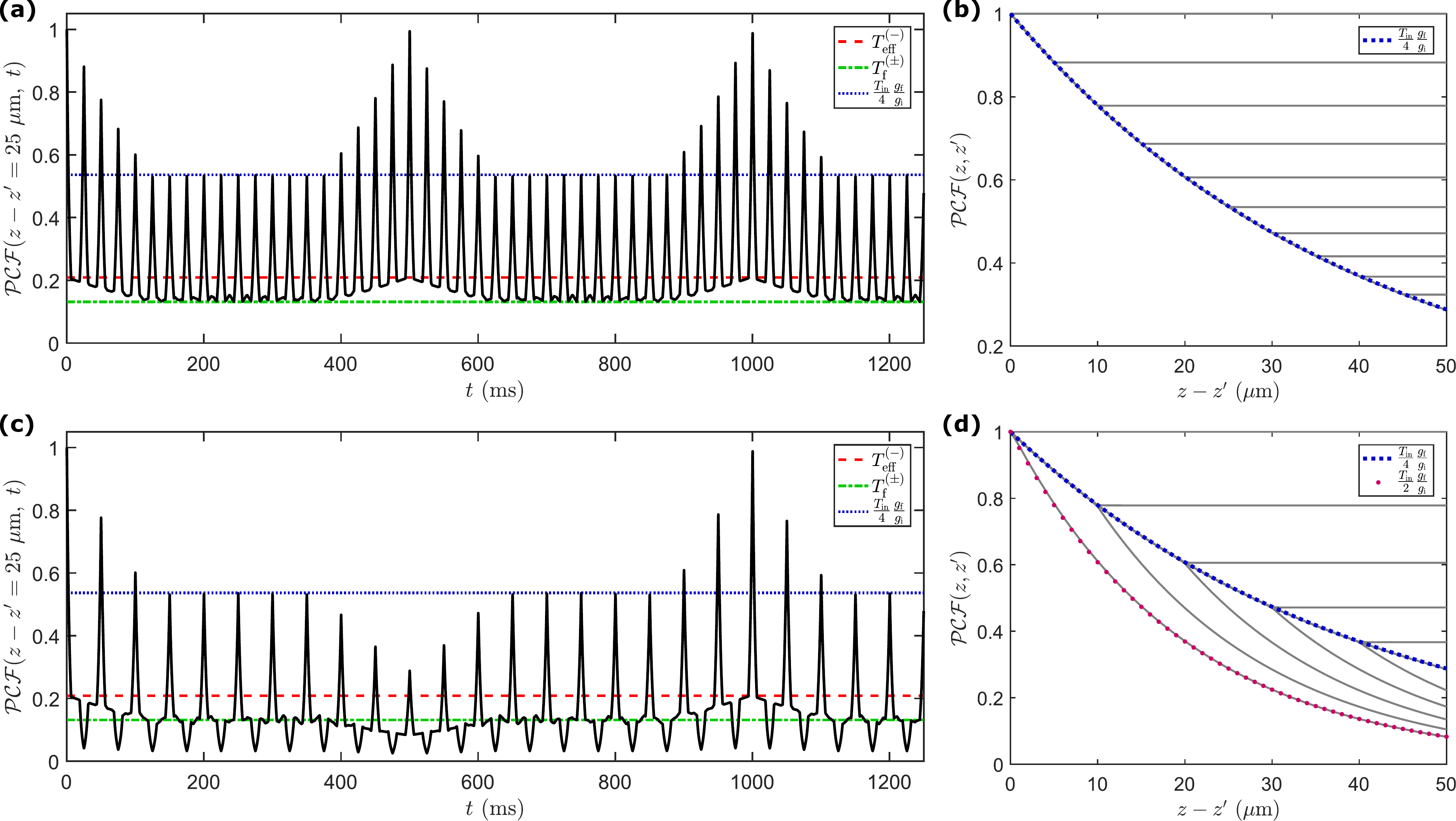}
	\caption{\textbf{Many-body revivals.} (a) Many-body revivals in a homogeneous system with $L=100\,\mu$m $\sim 465\times\xi_h$ and periodic boundary conditions. The initial temperature is $50\,$nK, the density $\rho_1+\rho_2=120\,/ \mu$m and $K=81$. All other parameters are the same as in Fig.~\ref{fig:figure2}. We observe fast and slow revivals which are connected with the different characteristic velocities $c_+$ and $c_-$. The corresponding revival times are  given by $t_-=n \times L/(2c_-)$ and $t_+=n \times L/(2c_+)$, with $n$ integer. The initial coherence is restored completely if fast and slow revivals coincide. Between the slow revivals, the fast revivals only restore coherence up to a value corresponding to a temperature $T_{in}/4\times g_\mathrm{f}/g_\mathrm{in}$ (cf. Eq.~\ref{eqn:phasevariance}). The minima of coherence between the fast revivals oscillate between $T_\mathrm{eff}^{(-)}$ and $T_f^{(\pm)}$. (b) By plotting  the $\mathcal{PCF}$  at the times of the fast revivals only, one directly observes a \textit{revival light-cone} with characteristic velocity $c_-$. Experimental box potentials are approximately described by fixed boundary conditions ($\delta \rho / \delta z = 0$, corresponding to a hard-walled box). In this case, the fast and slow revival times are doubled. Also, correlations are not translation invariant anymore. In (c) and (d) we plot the corresponding $\mathcal{PCF}$ and \textit{revival light-cone} for an ideal box trap of the same length as in (a). They exhibit essentially the same physics as in the case with periodic boundary conditions. Note the small, fast anti-revivals, which are a result of choosing the coordinates $z$ and $z^\prime$ symmetrically around the center of the trap.}
	\label{fig:figure4}
\end{figure*}

\section{INFLUENCE ON MANY-BODY REVIVALS}\label{sec:revivals}
While the double light-cone dynamics are clearly visible in the correlation functions calculated for an infinite system, observing them directly in an experiment with a finite size system, in particular when a typically harmonic longitudinal confinement is present, is challenging. In particular, due to the non-linear excitation spectrum in harmonic traps~\cite{Geiger13} the effect is severely scrambled by highly irregular many-body revivals. Examples of this behavior are shown in the Appendix. 

However, regular and well controlled many-body revivals have recently been observed for the first time in homogeneous trapping potentials~\cite{Rauer17}, demonstrating their power to probe the dephasing and higher-order interactions of phonon modes. In the following we thus illustrate the influence of our effect on such many-body revivals. 

To this end, we first repeat our calculations for a system with periodic boundary conditions, but with a typical experimental finite size of $100\,\mu$m $\sim 400 \times\xi_h$. In Fig.~\ref{fig:figure4}a we show the corresponding results for the phase correlation function. Due to the finite number of momentum modes they show clear rephasing behavior, as experimentally observed in Ref.~\cite{Rauer17}. However, due to the imbalance the two velocities in the system can be observed through the presence of two different types of revivals - slow revivals resulting from $c_+$ and fast revivals coming from $c_-$. The value of the phase correlation function reached in the slow revivals depends on how well slow and fast revivals coincide. 

A scenario more relevant for an experimental realization is the one of fixed boundary conditions ($\partial\phi/\partial z = 0$), which corresponds to the case of a hard walled box. This boundary conditions guarantee that the particle current at the box walls vanishes. The corresponding results are shown in Fig.~\ref{fig:figure4}b. Again, a clear distinction between slow and fast revivals can be observed, which can directly be connected to the two characteristic velocities. 

Our observations have important practical consequences for the experimental study of integrability-breaking in 1D Bose gases~\cite{Mazets08,Mazets10,Tan10,Burkov07,Stimming11,WeissPrivateComm,Tang17}. As the effect described here and true thermalization through integrability breaking would essentially lead to the same experimental signatures (i.e. thermal correlations corresponding to a temperature given by Eq.~\ref{eqn:temperature}) they would be very challenging to disentangle from measurements of correlation functions alone. In particular, any experimental effort clearly has to take both effects into account simultaneously. The many-body revivals presented in Fig.~\ref{fig:figure4} provide additional tools for such studies. 

\section{DISCUSSION}

We have observed how the dephasing of an imbalanced pair of 1D Bose gases can result in states which are, for all practical purposes, indistinguishable from thermal equilibrium. This is due to a coupling of the relative and common degrees of freedom that is mediated by the relative dephasing of the individual gases. It is important to note that this observation of an apparent thermalization relies on the thermal-like initial conditions that were imposed on the system by the coherent splitting process. The system always retains a strong memory of the initial conditions and thus has not truly reached global thermal equilibrium. For example, if the system was initialized with other non-thermal initial conditions like the ones demonstrated in~\cite{Langen2015}, it would equilibrate, but never appear thermal in its correlation functions~\cite{Linden09}. 

Interestingly, the observed dynamics are closely related to the measurement process. In experiments, fluctuations of the anti-symmetric degrees of freedom are probed. These degrees of freedom exhibit a rapid relaxation with a single time scale if there is no imbalance, and a relaxation with two distinct time scales if there is imbalance. The same timescales govern the relaxation of the symmetric degrees of freedom. In contrast to that, if the properties of a single gas were accessible in experiment, their individual correlations would already look completely relaxed after the first, rapid time scale. This highlights how even in integrable systems, observables need to be properly \textit{aligned} with (i.e. chosen such that they are sensitive to) the integrals of motion to reveal the integrable nature of the complex many-body dynamics. We note that the experiment in~\cite{Rauer17} recently revealed related behavior, where many-body revivals could be observed in certain correlation functions but not in others. This points to a general connection between the choice of measurement basis and the observed relaxation dynamics and will thus be an interesting topic for future research.

\bibliographystyle{apsrev4-1}
\bibliography{biblio}

\section*{ACKNOWLEDGEMENTS}
We acknowledge discussions with I. Mazets, T. Gasenzer, S. Erne and B. Rauer.  T.L. acknowledges support from the Alexander von Humboldt Foundation through a Feodor Lynen Fellowship, the EU through a Horizon2020 Marie Sk\l odowska-Curie IF (746525 coolDips), the Baden-W\"urttemberg Foundation and the Center for Integrated Quantum Science and Technology (IQST). T.S. acknowledges support by the Austrian Science Fund (FWF) through the Doctoral Programme CoQuS (\textit{W1210}). 
E.D.  acknowledges support from Harvard-MIT CUA, NSF Grant No. DMR-1308435, AFOSR Quantum Simulation MURI, ARO MURI on Atomtronics, ARO MURI Qusim program, and AFOSR MURI Photonic Quantum Matter. J.S. acknowledges support through the ERC advanced grant QuantumRelax. 

\cleardoublepage
\onecolumngrid
\section*{APPENDIX}

\subsection{Details of the proposed experimental realization}
The splitting quench can be realized by applying near-field RF radiation via two wires on an atom chip~\cite{Schumm05,Hofferberth06,Lesanovsky06}. Previous experiments investigating a balanced splitting process~\cite{Langen13,Langen13b,Gring12,Kuhnert13,Langen2015} have demonstrated this to be a powerful scenario for non-equilibrium physics. The splitting process can be made much faster than the speed of sound in the system, realizing the binomial distribution of atoms that is discussed in the main text. In this case, no information about the quench can propagate along the system, leading to almost perfectly correlated phase profiles of the two gases after the splitting~\cite{Langen13b}. 
The relative amplitude and phase of the two RF currents defines the polarization of the RF radiation and therefore the orientation of the double well potential~\cite{Schumm05,Hofferberth06}. A small imbalancing of in-phase RF currents in these wires creates a tilted double well potential and thus a small atomnumber imbalance after the splitting process. For the trap parameters used in the experiments we estimate that offsets below $250\,$Hz between the two minima of the tilted double well are sufficient to realize the scenario of small imbalances (i.e. up to $\Delta = 0.1$) discussed in the main text. In this case, the trapping potential provided by the two wells can still be assumed to be identical. Residual small collective excitations can be efficiently removed using optimal control~\cite{Mennemann15}. For higher imbalances, the wells become increasingly distorted until eventually also tunneling from one well across the barrier into excited states of the other well becomes possible. 

\subsection{Interference contrast}
\begin{figure*}[htb]
	\centering
		\includegraphics[width=0.60\textwidth]{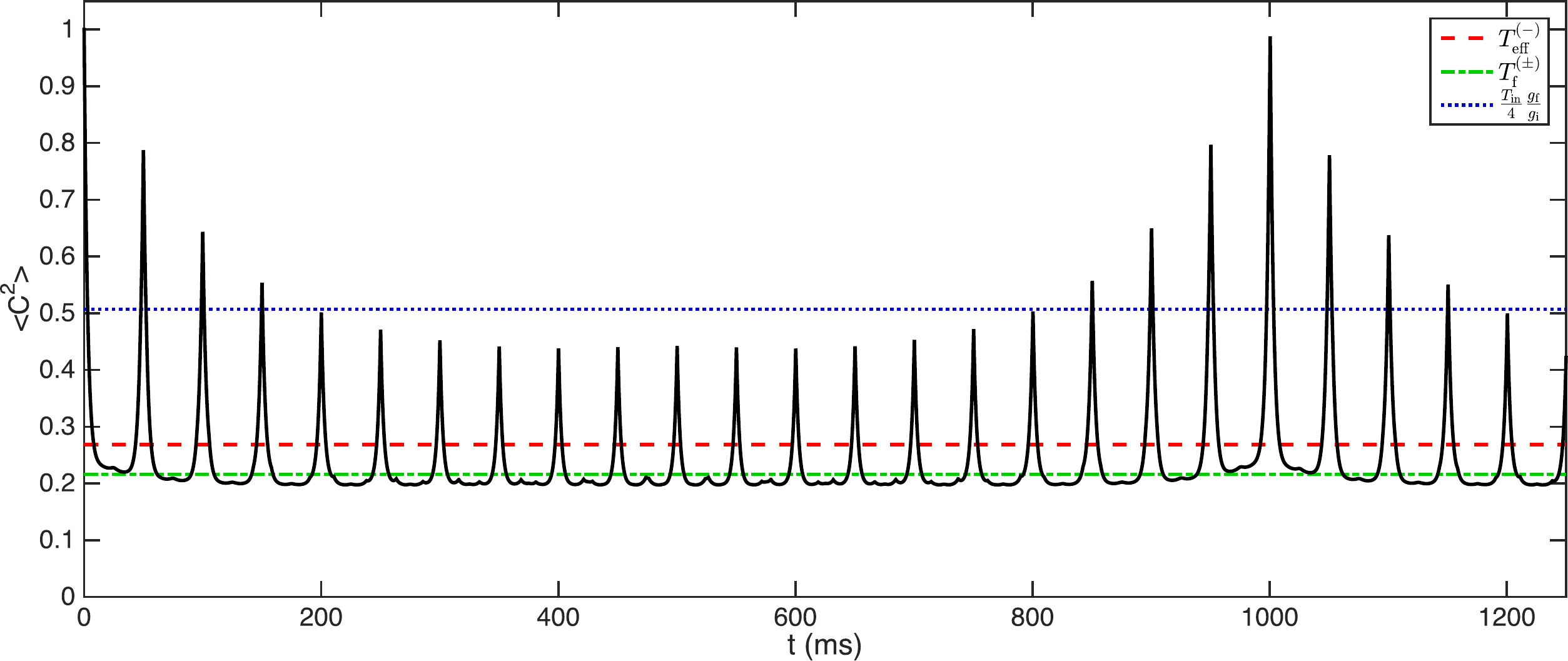}
	\caption{\textbf{Contrast as another experimental probe.} The slow and fast revivals that have been identified for the phase correlation function are also observable using the mean squared contrast $\langle C^2\rangle$. Here, we have used a system size of $100\,\mu$m, initial temperature of $T_\mathrm{in}=50\,$nK, imbalance $\Delta = 0.1$, trap frequencies $\omega_{\perp,\mathrm{in}}=2\pi\times2.1\,$kHz, $\omega_{\perp,\mathrm{f}}=2\pi\times1.4\,$kHz and initial density $\rho_1+\rho_2=120\,\mu\mathrm{m}^{-1}$.}
	\label{fig:figureA3}
\end{figure*}

In experiments the expectation value in the definition of the phase correlation function is realized through an average over many experimental runs. We have previously demonstrated~\cite{Langen13b} that the number of runs that is required for a statistically meaningful determination of the correlation functions can dramatically increase for evolution times $t\gg 10\,$ms. However, for the parameters used in this work, the fully thermal state is not expected before $t\sim 100\,$ms or more. For longer evolution times it could thus be beneficial to probe the level of coherence between the two gases using the mean squared interference contrast $\langle C^2\rangle$ of the matter wave interference pattern in time-of-flight~\cite{Gring12,Kuhnert13,Langen13b}. It is well known from one of our previous experiments~\citep{Kuhnert13} that this procedure involves an unknown factor that describes the finite resolution and other spurious experimental effects. We thus suggest to extract the relative phase $\theta(z)=\phi_L(z)-\phi_R(z)$ from every longitudinal position $z$ of the interference pattern and calculate the contrast via the identity $\langle C^2(L)\rangle=\int_{-L/2}^{L/2} \mathcal{PCF}(z_1,z_2) dz_1 dz_2$. Here, $\mathcal{PCF}$ is the two-point phase correlation function discussed in the main text (Eq.~\ref{eqn:PCF}) and $L$ is a length scale over which the interference pattern is integrated. Because of the integration this procedure can be more robust against statistical fluctuations than the phase correlation function alone. With the identity given above, it is straight forward to generalize our predictions for the dynamics to the contrast. An example is shown in Fig.~\ref{fig:figureA3}.

\subsection{Evolution of the phase variance}
In analogy to the phase variance for the anti-symmetric or relative phase in Eq.~\ref{eqn:phasevariance}, we obtain the following result for the phase variance of the symmetric degrees of freedom:
\begin{align}
\langle |\theta_k|^2\rangle&= \frac{m k_B T_\mathrm{in}}{\hbar^2 k^2 (\rho_1 + \rho_2)}\cos^2(c_- kt) \nonumber\\&\times \left[2+\frac{g_\mathrm{f}}{g_\mathrm{in}}+\left(2-\frac{g_\mathrm{f}}{g_\mathrm{in}}\right)\cos(2c_+kt)\right]\nonumber\\
	&+\frac{2mg_\mathrm{f}}{\hbar^2k^2}\sin^2(c_-kt)\cos^2(c_+ kt).\label{eqn:phasevariance2}
\end{align}
This represents the approximation for small imbalances, as does Eq.~\ref{eqn:phasevariance} for the relative phase.

Similarly to Eq.~\ref{eqn:singlegases_phvar_evo} for the phase variances of the individual condensates, one can also calculate the evolution of the cross-terms $\langle \hat{\phi}_{1,k} \ \hat{\phi}_{2,-k}  \rangle$. From Eq.~\ref{eq:operator_evolution} and \ref{eq:mix_init} it is easy to see that they are of the form
\begin{align}
\begin{split}
\langle \hat{\phi}_{1,k} \ &\hat{\phi}_{2,-k}  \rangle  \\ 
=  \ &\mathcal{C}_1 \cdot \cos(c_1kt) \cos(c_2kt) + \mathcal{C}_2 \cdot \sin(c_1kt) \sin(c_2kt) \\
 = \  &\frac{\mathcal{C}_1}{2} \cdot \left(\cos(2 c_-kt) + \cos(2 c_+kt) \right) + \\
   &\frac{\mathcal{C}_2}{2} \cdot \left(\cos(2 c_-kt) - \cos(2 c_+kt) \right),
\end{split}
\end{align}
where $\mathcal{C}_{1,2}$ are time-independent constants. This expression describes a dephasing of the cross terms to $0$ with the velocity difference $c_-$.

\subsection{Harmonic traps}
The Luttinger Hamiltonian as written in Eq.~\ref{eqn:llhamiltonian} stays valid for inhomogeneous density profiles $\rho_i(z)$. For the calculation in the harmonic trap, we assume a Thomas-Fermi profile for the density distribution before splitting, and rescaled density distributions for the evolution of the fluctuations after the quench. For the latter, the initial density profile depending on the total atomnumber $N_1 + N_2$ and on the trap frequencies $\omega_z$ (longitudinal) and $\omega_{\perp,\mathrm{in}}$ (radial) is simply multiplied by the factor $N_i/(N_1 + N_2)$. 

With this density distributions the Hamiltonian can be diagonalized with the help of Legendre-Polynomials~\cite{Geiger13,Petrov04}. Note that due to the density dependence of the shot-noise fluctuations introduced in the splitting process, the initial density fluctuations expanded in Legendre-Polynomials are not diagonal anymore~\cite{ErnePrivComm}.

\begin{figure*}[htb]
	\centering
	\includegraphics[width=1\textwidth]{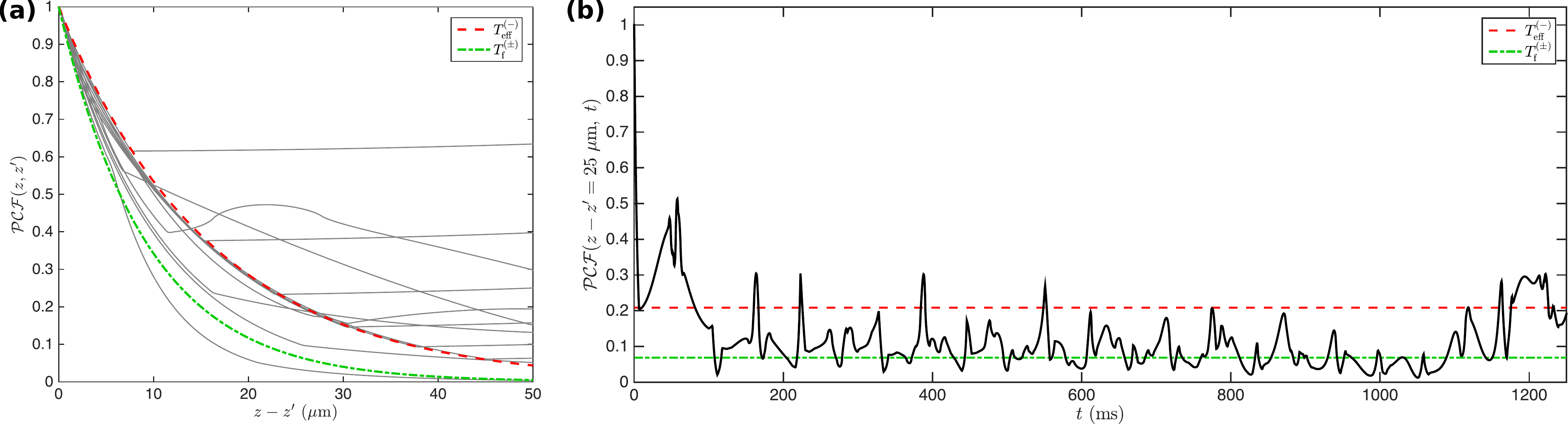}
	\caption{\textbf{Dynamics in a harmonic trapping potential.} (a) Dynamics of the phase correlation function, with time increasing in steps of $2\,$ms from $0\,$ms to $12\,$ms, and in steps of $25\,$ms from $35\,$ms to $160\,$ms. (b) The incommensurate mode energies of the trapped system lead to complex dephasing dynamics with many irregularly spaced partial revivals. We have used $10000$ ${}^{87}$Rb atoms with an initial temperature of $T_\mathrm{in}=70\,$nK, in an initial trap with frequencies $\omega_z=2\pi\times11\,$Hz and $\omega_{\perp,\mathrm{in}}=2\pi\times2.1\,$kHz. The atomnumber-imbalance for the evolution is $\Delta = (N_1 - N_2)/(N_1 + N_2) = 0.1$, the perpendicular trap frequency is $\omega_{\perp,\mathrm{f}}=2\pi\times1.4\,$kHz. Note that we don't need to specify a longitudinal trap frequency for the evolution after the quench as we simply assume rescaled density profiles.}
	\label{fig:figureA1}
\end{figure*}

The results of the calculation are shown in Fig~\ref{fig:figureA1}. The incommensurate excitation energies of the trapped system lead to very complex dephasing and rephasing dynamics. As already discussed in the main text, such complex dynamics make it very challenging to experimentally disentangle different competing integrable (such as the one presented here) and non-integrable (such as thermalization) mechanisms. 

\subsection{Spin-charge coupling in 1D Bose gases}
The experimental realization of our scenario can also be interpreted as a platform to explore spin-charge physics within the Luttinger liquid framework~\cite{Giamarchi04,Schmidt10}. In this case, the symmetric degrees of freedom can be identified with the charge degrees of freedom of a fermionic spin chain, while the anti-symmetric degrees of freedom play the role of the spin. If the two gases are prepared with identical mean atom numbers, spin and charge degrees of freedom are separated. The mixing in the imbalanced case, on the other hand, can be identified as a coupling between spin and charge. 

For the system of two spatially separate 1D Bose gases the characteristic velocities $c_{s,c}=\sqrt{g_{s,c}\rho/m}$ of spin and charge degrees of freedom are identical, as $g_s=g_c\equiv g$, where $g$ is the 1D interation strength. Different tunable velocities for spin and charge can be achieved by replacing the two wells employed in this work by two internal atomic states $|1\rangle$ and $|2\rangle$ with different interaction constants $g_{11}$, $g_{22}$ and $g_{12}$~\cite{Kitagawa11,Widera08,Zvonarev07,Fuchs05}. This situation would lead to $g_{s,c} = g_{11}+g_{22}\mp 2 g_{12}$ and thus different velocities for spin and charge. These velocities could be studied experimentally by probing the propagation of the \textit{in situ} density fluctuations after a quench of the radial confinement~\cite{Esteve06,Hung13}.

\end{document}